\documentclass[aps,prl,reprint,showpacs,superscriptaddress,twocolumn]{revtex4-1}

\usepackage[english]{babel}
\usepackage{amsmath,bm}
\usepackage{amsmath,amssymb,bbm,graphicx,comment,dsfont,braket,mathtools,units}
\usepackage{hyperref}
\hypersetup{colorlinks=true,citecolor=blue,linkcolor=blue,urlcolor=blue,pdfstartview=FitH}

\usepackage{color}
\definecolor{dkgreen}{rgb}{0,0.5,0}

\newcommand{\norder}[1]{ {\mkern1mu\colon\mkern-4mu{#1}\colon\mkern-3mu} }
\newcommand{\abs}[1]{\left\lvert #1 \right\rvert}
\newcommand\dif{\mathop{}\!\mathrm{d}}
\newcommand{\be}{\begin{equation}}
\newcommand{\ee}{\end{equation}}
\newcommand{\up}{\mathord{\uparrow}}
\newcommand{\dn}{\mathord{\downarrow}}
\newcommand{\RePart}{\operatorname{Re}}
\newcommand{\ImPart}{\operatorname{Im}}
\DeclareMathOperator{\erfc}{erfc}
\DeclareMathOperator{\erf}{erf}

\begin{document}

\date{\today}

\title{Time-domain anyon interferometry in Kitaev honeycomb spin liquids and beyond}
\author{Kai Klocke}
\affiliation{Department of Physics and Institute for Quantum Information and Matter, California Institute of Technology, Pasadena, CA 91125, USA}
\affiliation{Department of Physics, University of California, Berkeley, California 94720, USA}
\author{David Aasen}
\affiliation{Kavli Institute for Theoretical Physics, University of California, Santa Barbara, California 93106, USA}
\affiliation{Microsoft Quantum, Microsoft Station Q, University of California, Santa Barbara, California 93106-6105, USA}
\author{Roger S. K. Mong}
\affiliation{Department of Physics and Astronomy, University of Pittsburgh, Pittsburgh, PA 15260, USA}
\affiliation{Pittsburgh Quantum Institute, Pittsburgh, PA 15260, USA}
\author{Eugene A. Demler}
\affiliation{Department of Physics, Harvard University, 17 Oxford st., Cambridge, MA 02138, USA}
\author{Jason Alicea}
\affiliation{Department of Physics and Institute for Quantum Information and Matter, California Institute of Technology, Pasadena, CA 91125, USA}
\affiliation{Walker Burke Institute for Theoretical Physics, California Institute of Technology, Pasadena, CA 91125, USA}

\begin{abstract}
Motivated by recent experiments on the Kitaev honeycomb magnet $\alpha\text{-RuCl}_3$, we introduce time-domain probes of the edge and quasiparticle content of non-Abelian spin liquids.  
Our scheme exploits ancillary quantum spins that communicate via time-dependent tunneling of energy into and out of the spin liquid's chiral Majorana edge state.  
We show that the ancillary-spin dynamics reveals the edge-state velocity and, in suitable geometries, detects individual non-Abelian anyons and emergent fermions via a time-domain counterpart of quantum-Hall anyon interferometry.
We anticipate applications to a wide variety of topological phases in solid-state and cold-atoms settings.  
\end{abstract}

\maketitle

{\bf \emph{Introduction.}}~Topologically ordered phases that support non-Abelian anyons---fractionalized quasiparticles exhibiting non-commutative braiding statistics---provide a potential quantum-computing medium with intrinsic fault-tolerance \cite{Kitaev:2003,TQCreview}.
To this end, developing single-anyon-detection techniques poses a key challenge, both for validating anyonic content and for readout.  
Earliest efforts centered around the fractional quantum Hall (FQH) state at filling $\nu = 5/2$ \cite{eisenstein_1987}, which is now widely believed to realize the non-Abelian Moore-Read state \cite{MooreRead} (or one of its cousins \cite{Levin2007,Lee2007,SonDiracCFL}) \cite{Banerjee}.
There, electrical anyon interferometry \cite{Nayak2005, Stern2006, Bonderson2006, Bishara2008} enables single-anyon detection 
and has been explored both at $\nu = 5/2$ \cite{Willett2019} and especially decisively in the Abelian $\nu = 1/3$ state \cite{manfra_2020}.  

More recent experiments spotlight a fundamentally different candidate non-Abelian-anyon platform: quantum spin liquids in spin-orbit-coupled Mott insulators governed by variants of Kitaev's honeycomb model \cite{Kitaev2006, Jackeli_2009,Trebst2017}.  
In particular, thermal-transport measurements \cite{Kasahara2018, Tokoi} on $\alpha\text{-RuCl}_3$ \cite{Plumb2014} suggest that the non-Abelian spin liquid phase from Kitaev's model emerges upon application of an $\mathcal{O}(10$T) magnetic field (see also Ref.~\onlinecite{chern2020sign}).  
This putative spin liquid mimics the Moore-Read state modulo the charge sector; it supports a chiral Majorana edge state, trivial bosonic excitations ($\mathds{1}$), emergent fermions ($\psi$), and `Ising' non-Abelian anyons ($\sigma$).  
Crucially, however, detection methods do not directly carry over from the quantum Hall problem due to the host system's Mott-insulating character.
While several recent works have nevertheless devised electrical spin-liquid probes \cite{aasen2020,zhang2020electrical,feldmeier2020local,knig2020tunneling,pereira2020electrical,udagawa2020stm}, the problem certainly warrants renewed attention.  

\begin{figure}
	\centering
	\includegraphics[width=.99\columnwidth]{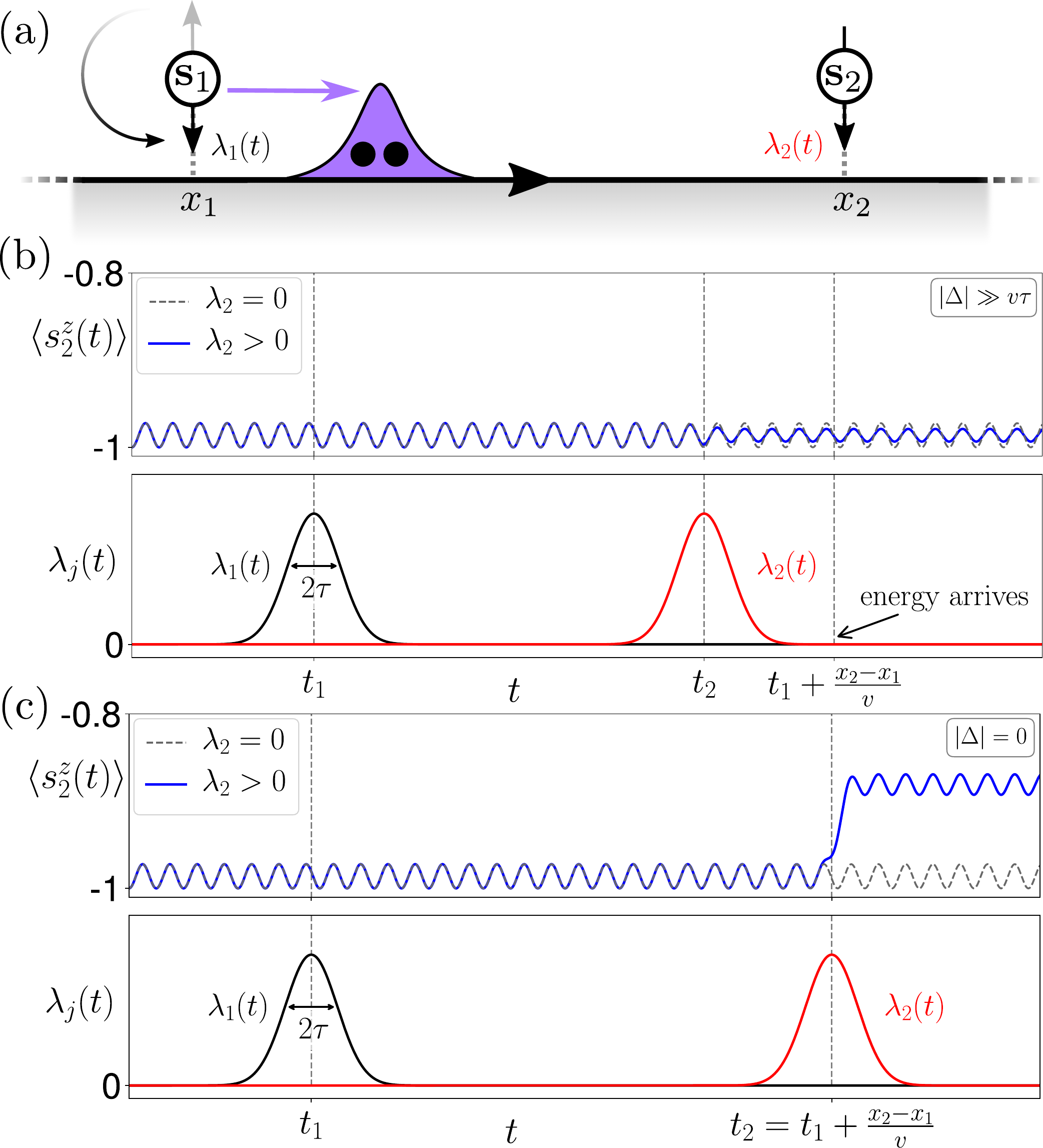}
	\caption{(a) Ancillary spins ${\bf s}_{1,2}$ interacting with a chiral Majorana edge state via time-dependent couplings $\lambda_{1,2}(t)$.  
	At time $t = 0$, ${\bf s}_1$ and ${\bf s}_2$ are respectively initialized into excited (up) and ground-state (down) spin configurations.	
	Pulsing $\lambda_1(t)$ allows ${\bf s}_1$ to relax, depositing excess energy (purple) into the edge as a pair of Majorana fermions (black dots) that propagate chirally toward ${\bf s}_2$.   
	An `aligned' $\lambda_2(t)$ pulse, timed to coincide with the arrival of the energy packet, allows ${\bf s}_2$ to retrieve the incident energy.  	(b) Time evolution of $\langle s_2^z(t)\rangle$ (top panel) for a `misaligned' pulse sequence (bottom panel) such that $\lambda_2(t)$ turns on prior to the arrival of energy injected by ${\bf s}_1$.  
	The small suppression of oscillations after the misaligned $\lambda_2$ pulse reflects an interplay with spin precession.
	(c) Same as (b), but for an aligned $\lambda_2(t)$ pulse.  
	Energy shuttling from ${\bf s}_1$ to ${\bf s}_2$ mediates a spin flip responsible for the late-time `kick' in $\langle s^z_2(t)\rangle$.
	Parameters for (b,c): $h_x / h_z = 1/6$, $h_z \tau = 3.$, $\tau / (t_2 - t_1) = 1/20$, $\bar\lambda_{1,2}h_z/v^2 = 1.275$.
	}
	\label{fig:unpinchedGeometry}
\end{figure}

\begin{figure*}
	\centering
	\includegraphics[width=.99\textwidth]{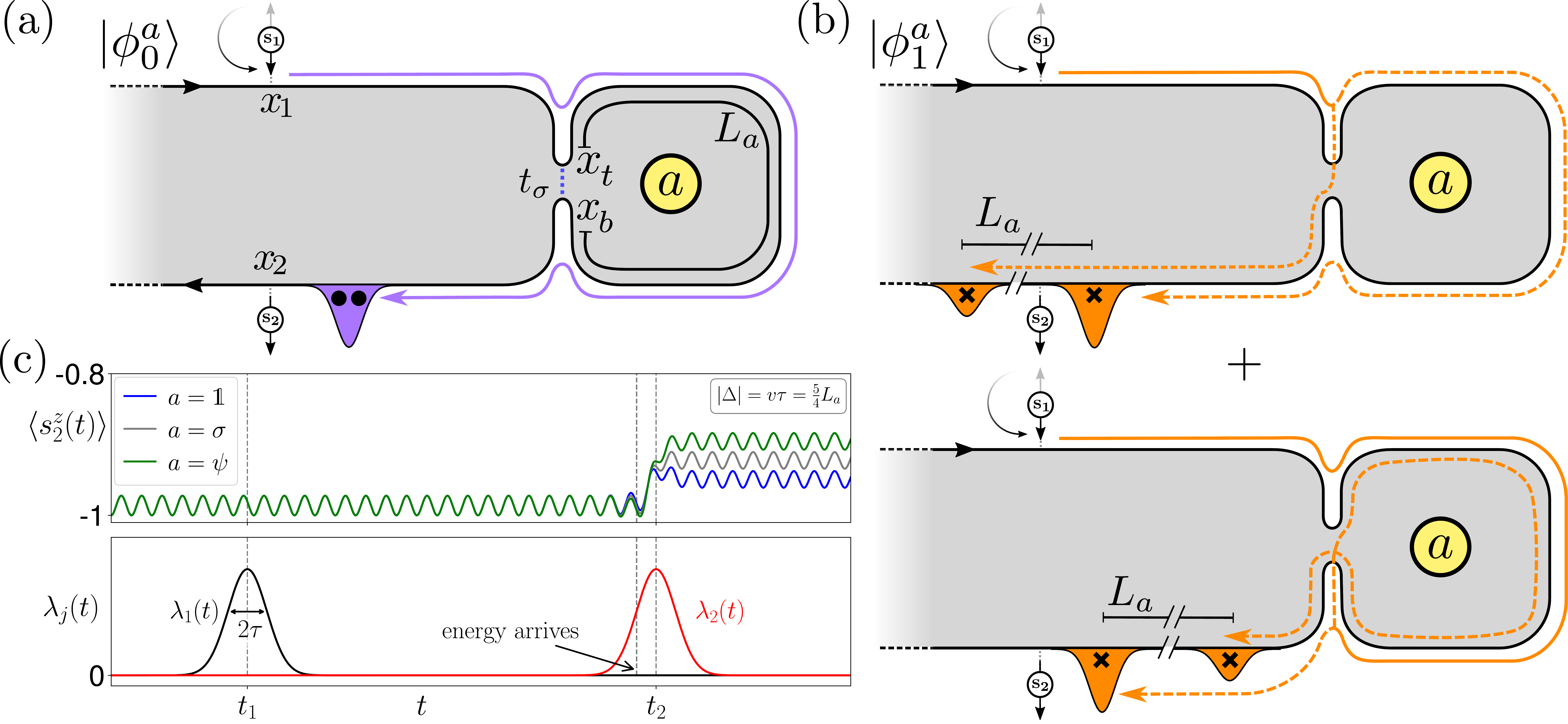}
	\caption{(a,b) Interferometer enabling detection of a bulk quasiparticle of type $a = \mathds{1},\psi,$ or $\sigma$ residing in the enclosed region of length $L_a$.	
	The initialization and pulse sequence are the same as in Fig.~\ref{fig:unpinchedGeometry}.  
	For the trivial path depicted in (a), ${\bf s}_1$ injects energy that bypasses the constriction and travels the long way toward ${\bf s}_2$.
	For the two $\mathcal{O}(t_\sigma)$ tunneling paths in (b), the energy instead splinters into Ising anyons (indicated by x's) at the constriction, one of which encircles quasiparticle $a$; the wavepackets for the outgoing Ising-anyon edge excitation are separated by $L_a$.	
	Interference between these paths yields an $a$-dependent probability for energy absorption by ${\bf s}_2$.
	(c) Time evolution of $\langle s^z_2(t)\rangle$ (top panel) for a partially misaligned ($\abs{\Delta} = \frac54 L_a$) $\lambda_{1,2}(t)$ pulse sequence (bottom panel).
	Crucially, the late-time behavior discriminates the three quasiparticle types.
	Parameters are $\tilde{t}_\sigma = 2$ and $v\tau / L_a = 5/4$, with others as given in Fig.~\ref{fig:unpinchedGeometry}(c).		
	}
	\label{fig:pinchedGeometry}
\end{figure*}

We introduce a single-anyon detection scheme naturally tailored to a Mott-insulating spin system.
Our approach eschews electrical measurements altogether in favor of time-domain probes of ancillary spins dynamically coupled to the spin-liquid's chiral Majorana edge state.
Time-domain techniques have been profitably employed to study chiral topological phases in various contexts, including detection of edge magnetoplasmons \cite{Ashoori_1992}, generation of coherent single-electron excitations on quantum Hall edges \cite{Keeling_2008}, and edge-mediated state transfer \cite{Yao_2013}.
In our proposal, an `emitter' ancillary spin shuttles (bosonic) energy via the chiral Majorana edge state towards a downstream `absorber' ancillary spin [Fig.~\ref{fig:unpinchedGeometry}(a)].  
If the spin liquid contains a constriction, as in Fig.~\ref{fig:pinchedGeometry}, en route the injected energy can splinter such that a fractionalized edge excitation encircles a bulk quasiparticle (of type $\mathds{1}, \psi,$ or $\sigma$).  
Crucially, the probability of energy capture by the absorber spin depends on the bulk quasiparticle type by virtue of nontrivial braiding statistics---enabling single-anyon detection via a time-domain analogue of FQH interferometry.
This scheme extends to general Abelian and non-Abelian chiral topological phases and appears particularly well-suited for insulating magnets and cold atoms.

{\bf \emph{Edge-state interrogation.}}~We first illustrate how our methods enable time-domain exploration of edge states.  Suppose that two ancillary spin-1/2 degrees of freedom ${\bm{s}}_{1,2}$ locally couple to the spin liquid's chiral Majorana edge mode at positions $x_{1,2}$ (Fig.~\ref{fig:unpinchedGeometry}a).
We model the dynamics with a Hamiltonian \cite{affleck_2020}
\be
	H = -iv\int_x \gamma \partial_x \gamma + \sum_{j=1,2}\left[\mathbf{h}\cdot\bm{s}_j + \frac{\lambda_j(t)}{2\pi} s_j^x T(x_j) \right],
	\label{eq:baseHamiltonian}
\ee
where the Majorana field obeys $\{\gamma(x),\gamma(x')\} = \frac12\delta(x-x')$, $s_j^\alpha$ are Pauli operators acting on the ancillary spins, and $T(x) = -2\pi i \norder{\gamma\partial_x\gamma}$ is the normalized stress-energy tensor for the edge conformal field theory (CFT).
The first term in Eq.~\eqref{eq:baseHamiltonian} describes the edge kinetic energy with velocity $v$ while the second captures the ancillary-spin Zeeman energy.
Throughout we assume for simplicity $h_z \gg \lvert h_x \rvert$ and $h_y = 0$.
The third hybridizes the ancillary spins to the edge state via couplings $\lambda_j(t)$ that descend from exchange interactions with the non-Abelian spin liquid \cite{aasen2020,Yao_2013};   
these terms mediate energy shuttling between the ancillary spins by allowing each spin to locally absorb or deposit energy packets consisting of an \emph{even} number of fermionic edge excitations.
We assume that the $\lambda_j(t)$ couplings, and hence shuttling of energy, are amenable to real-time control.  

Figure~\ref{fig:unpinchedGeometry} sketches the protocol of interest: 
$(i)$ Start with $\lambda_{1,2} = 0$ and prepare an initial state $\ket{\phi(t = 0)} = \ket{0} \otimes \ket{s^z_1 = \up, s^z_2 = \dn}$, where $\ket{0}$ describes the vacuum for the Majorana edge mode; note the excess energy $\sim 2h_z$ for spin $\bm{s}_1$.
 $(ii)$ Turn on a Gaussian pulse $\lambda_1(t) = \bar\lambda_1 e^{-(t-t_1)^2/(2\tau^2)}$ that enables $\bm{s}_1$ to shed energy into the edge, where it propagates chirally toward $\bm{s}_2$ at speed $v$.  
$(iii)$ Turn on $\lambda_2(t) = \bar\lambda_2 e^{-(t-t_2)^2/(2\tau^2)}$.
We take $h_z \tau \gg 1$ so that the pulses approximately conserve energy \footnote{Any pulse shape $\lambda_j(t)$ suffices provided it becomes negligible after some finite time and varies sufficiently smoothly.}. 
If the duration between pulses satisfies $v(t_2 - t_1) \approx x_2 - x_1$ (within a tolerance of $v\tau$), then energy that $\bm{s}_1$ deposits to the edge arrives coincident with the $\lambda_2$ pulse and can thus be absorbed by $\bm{s}_2$.
Spin $s^z_2$ is measured at some time $t>0$.  

We compute the expectation value $\langle s^z_2(t)\rangle$ perturbatively in 
$h_x/h_z$ and $\Lambda_j \equiv \bar \lambda_j h_z/v^2$ (the dimensionless spin-edge coupling strength at energy scale $h_z$), assuming $(t_2 - t_1) \gg \tau$.
For details see Appendix~\ref{app:PerturbS2Z}.
At measurement times $t \gg t_2$ we find
\be
\begin{aligned}
\langle s_2^z(t) \rangle &\approx -1 + \left(h_x/h_z\right)^2 \sin^2(h_z t)
\\
& -\frac{1}{12\sqrt{\pi}} \Lambda_2^2\left(\frac{h_x}{h_z}\right)^2(h_z\tau)\cos(2h_z t)
\\
& + \frac{1}{36\pi}\left(\Lambda_1\Lambda_2\right)^2(h_z \tau)^2 e^{-\frac{\Delta^2}{2(v\tau)^2}},
\end{aligned}\label{eq:sz2_expectation}
\ee
where $\Delta = v(t_2 - t_1) - (x_2-x_1)$ quantifies the timing mismatch between the pulses.
The first line reflects spin precession from the Zeeman field.
The second line is independent of pulse timing and originates from processes whereby $h_x$ flips $s_2^z$ from down to up, after which $\lambda_2$ mediates a second spin flip.
Most importantly, the final line---which depends exponentially on pulse misalignment $\Delta$---is the correction due to energy shuttling from $\bm{s}_1$ to $\bm{s}_2$.  
Figures~\ref{fig:unpinchedGeometry}(b) and (c) respectively display $\langle s_2^z(t) \rangle$ for misaligned ($\abs{\Delta} \gg v\tau$) and aligned ($\abs{\Delta} \approx 0$) pulses.
Observing the additional `kick' visible in the aligned case provides a direct measure of the edge-state velocity $v$.  
If phase coherence is maintained over the length $x_2 - x_1$ then we recover an additional oscillatory correction due to interference between the energy shuttling and the precession of both spins (see Appendix~\ref{app:PerturbS2Z}).

{\bf \emph{Time-domain anyon interferometry.}}~Next we revisit the above double-pulse protocol in an interferometer geometry featuring a constriction in the spin liquid [Figs.~\ref{fig:pinchedGeometry}(a,b)].  
We are specifically interested in the energy-shuttling probability when a bulk quasiparticle of type $a = \mathds{1}, \psi,$ or $\sigma$ resides in the enclosed region.
At the constriction, bosonic energy packets injected from spin ${\bf s}_1$ can splinter into fractionalized edge excitations.
We include only the most relevant process wherein Ising anyons tunnel between positions $x_t$ and $x_b$ across the pinch \cite{Fendley2009,Fendley2007,Fendley_2006}:
\be
	H_\text{tun} = t_\sigma e^{-i\pi h_\sigma}\sigma(x_b)\sigma(x_t).
	\label{eq:tunnelHamiltonian}
\ee
Here $\sigma(x)$ is the Ising-anyon field with conformal weight $h_\sigma = 1/16$ and $x_b - x_t \equiv L_a$ defines the path length enclosing quasiparticle $a$.  
We work in the regime where the dimensionless tunneling parameter $\tilde t_\sigma \equiv t_\sigma L_a^{7/8} / v$ admits a perturbative treatment \cite{aasen2020}.

Let $\ket{\phi^a(t)} = \ket{\phi^a_0(t)} + \ket{\phi^a_1(t)} + \cdots$ denote the system's wavefunction with $\ket{\phi^a_j(t)}$ the $\mathcal{O}(\tilde t_\sigma ^j)$ component.  
To $\mathcal{O}(\tilde t_\sigma)$ we have $\langle s^z_2(t) \rangle = f_0(t) + f^a_{\rm int}(t)$, where
\begin{align}
  f_0 = \bra{\phi^a_0}s^z_2\ket{\phi^a_0}, ~~~f^a_{\rm int} = 2 \RePart\bra{\phi^a_0}s^z_2\ket{\phi^a_1}.
  \label{f0fint}
\end{align}
In the dominant wavefunction component, $\ket{\phi^a_0}$, energy deposited by ${\bf s}_1$ travels a distance $x_2-x_1$ the long way around the constriction before reaching ${\bf s}_2$.
For $h_x \not=0$ there is also a term in $\ket{\phi^a_0}$ involving precession of both spins.
The associated contribution $f_0(t)$ to $\langle s^z_2(t) \rangle$ from this trivial path is given at late times by Eq.~\eqref{eq:sz2_expectation}.
In the subleading wavefunction component, $\ket{\phi^a_1}$, energy imparted by ${\bf s}_1$ splinters into two Ising anyons---one of which carries some fraction of the incident energy across the constriction and, crucially, encircles quasiparticle $a$.
Two such paths arise [see Fig.~\ref{fig:pinchedGeometry}(b)] depending on whether tunneling happens from above or below the constriction.  
Energy packets that hop across the constriction travel a distance $(x_2 - x_1) \pm L_a$, while the remainder of the energy travels a distance $x_2-x_1$, coincident with the trivial path.  [We assume that ${\bf s}_2$ sits sufficiently far from the constriction that the tunneled Ising anyon from Fig.~\ref{fig:pinchedGeometry}(b), bottom, completely braids around $a$ before any energy arrives to $x_2$.]
Interference between the trivial path and the two $\mathcal{O}(\tilde t_\sigma)$ paths depends on the enclosed quasiparticle type $a$ and underlies the correction $f^a_{\rm int}$ in Eq.~\eqref{f0fint}.

Non-Abelian statistics kills interference for $a = \sigma$, i.e., $f_{\rm int}^{a = \sigma} = 0$.
Indeed, the braiding process where the edge Ising anyon encircles the bulk Ising anyon nontrivially rotates the system's quantum state such that $\ket{\phi^a_0}$ and $\ket{\phi^a_1}$ become orthogonal (similar to electrical non-Abelian FQH interferometry \cite{Stern2006, Bonderson2006}).  
Interference can, however, survive for $a = \mathds{1},\psi$---provided ${\bf s}_2$ also retrieves energy $\sim 2h_z$ in the paths from Figs.~\ref{fig:pinchedGeometry}(b). 
In these $\mathcal{O}(\tilde t_\sigma)$ paths, energy partitions between the edge Ising anyons in all ways compatible with energy conservation (see Appendix~\ref{app:IsingTunnel}), ensuring a finite energy-retrieval probability even when the Ising-anyon wavepackets in Fig.~\ref{fig:pinchedGeometry}(b) are well-separated.  
The interference corrections for $a = \mathds{1}, \psi$ satisfy $f_{\rm int}^{a = \psi} = -f_{\rm int}^{a = \mathds{1}}$, where the minus sign reflects the Abelian statistical phase acquired when the edge Ising anyon encircles a bulk fermion.
Thus $s^z_2$ measurements distinguish all three bulk quasiparticle types as summarized in Fig.~\ref{fig:pinchedGeometry}(c).

For a quantitative treatment, we take $a = \mathds{1}$ and evaluate the $\mathcal{O}(\Lambda_1^2\Lambda_2^2 \tilde t_\sigma)$ energy-shuttling interference correction to $\langle s_2^z(t) \rangle$ at late times $t\gg t_2$.
The trivial path corresponds to 
\begin{align}
  \ket{\phi_0^{a = \mathds{1}}(t \gg t_2)} & \approx - e^{i\frac{2h_z}{v} (x_2-x_1)}\frac{1}{6\sqrt{\pi}}\Lambda_1\Lambda_2(h_z \tau)
  \nonumber \\
  &\times \ket{0}\otimes\ket{s_1^z = \dn, s_2^z = \up} + \cdots, 
\end{align}  
where we displayed only the term describing energy shuttling from ${\bf s}_1$ to ${\bf s}_2$ that is relevant for interference.  
Overlap with the nontrivial energy-shuttling paths encoded by $\ket{\phi_1^{a = \mathds{1}}(t \gg t_2)}$ thus follows from
\be
\begin{aligned}
& \bra{0}\otimes\braket{\dn,\up \vert \phi_1^{a = \mathds{1}}} =  i \int\limits_{t_a, t_b, t_c} \frac{\lambda_1(t_a) \lambda_2(t_b)}{(2\pi)^2} e^{2ih_z(t_b-t_a)} \\
& \times \bra{0} T(t_b, x_2) \left[H_\text{tun}(t_c) - \langle H_\text{tun}(t_c) \rangle \right]T(t_a, x_1) \ket{0}.
\end{aligned}
\label{eq:fInt_setup}
\ee
The $\langle H_{\rm tun} \rangle$ term simply compensates the correction to the vacuum state due to Ising-anyon tunneling.
Appendix~\ref{app:IsingTunnel} evaluates the CFT correlator in Eq.~\eqref{eq:fInt_setup}, and in the late-time limit obtains
\begin{align}
f_{\rm int}^{a = \mathds{1}}(t\gg t_2) &\approx -\frac{\tilde t_\sigma}{192} \left(\Lambda_1 \Lambda_2\right)^2\left(\frac{v\tau}{L_a}\right)\left(h_z\tau\right)  \sin(2h_z L_a) 
\nonumber \\ 
&\times e^{-\frac{\Delta^2}{(2v\tau)^2}} \left[e^{-\frac{(\Delta + L_a)^2}{(2v\tau)^2}} - e^{-\frac{(\Delta - L_a)^2}{(2v\tau)^2}}\right].
\label{eq:Ptun}
\end{align}
Appendix~\ref{app:IsingTunnel} further derives the correction from interference between the $\mathcal{O}(\tilde t_\sigma)$ energy-shuttling paths and the precession of both spins---which exhibits oscillatory dependence on the measurement time $t$.

Several comments are warranted.  
After energy retrieval by ${\bf s}_2$, the edge returns to the vacuum state in the Fig.~\ref{fig:pinchedGeometry}(a) path but retains two Ising-anyon wavepackets separated by a distance $L_a$ in the $\mathcal{O}(\tilde t_\sigma)$ Fig.~\ref{fig:pinchedGeometry}(b) paths.  
Consequently, the late-time interference correction is proportional to $\langle 0| \sigma(x) \sigma(x+L_a)|0\rangle \sim L_a^{-1/8} \sim \tilde t_\sigma/L_a$, explaining the power-law $L_a$ dependence in Eq.~\eqref{eq:Ptun}.
The two bracketed exponentials in Eq.~\eqref{eq:Ptun} correspond to the processes where ${\bf s}_2$ absorbs energy $\sim 2h_z$ from the \emph{tunneled} Ising anyon in Fig.~\ref{fig:pinchedGeometry}(b).
These paths accumulate a dynamical phase $\pm 2h_z L_a$ relative to Fig.~\ref{fig:pinchedGeometry}(a) due to the path-length difference---hence the $\sin$ factor in Eq.~\eqref{eq:Ptun}.
[The Ising-anyon energy packet that remains coincident with the trivial path acquires no relative phase, and thus its absorption by ${\bf s}_2$ does not contribute to interference between Figs.~\ref{fig:pinchedGeometry}(a) and (b).]
For wide pulses with $v \tau \gtrsim L_a$, $\lambda_2(t)$ temporally overlaps with both edge Ising anyons and enables ${\bf s}_2$ to draw energy from both the advanced and delayed packets. 
The resulting interference signal is maximized when the pulse width and interference path length are comparable $(v\tau \sim L_a)$ and when the timing favors one of the tunneling paths $(\Delta \sim \pm L_a/2)$.
If the ancillary-spin Zeeman splitting can be controlled, the oscillatory $h_z$ dependence provides an additional interferometric signature.
For very narrow pulses such that $v\tau \ll L_a$, $\lambda_2(t)$ cannot have appreciable temporal overlap with the arrival of energy both from the trivial path and from one of the displaced Ising-anyon packets.
Under such conditions interference instead arises from the process where both spins precess in the Zeeman field, which allows for overlap between $\lambda_2(t)$ and the splintered energy packet for timing $\Delta = 0, \pm L_a$; see Appendix~\ref{app:IsingTunnel}.

Fermions can also tunnel across the constriction, adding a less-relevant $-i t_\gamma \gamma(x_b)\gamma(x_t)$ term to Eq.~\eqref{eq:tunnelHamiltonian}.
At $\mathcal{O}(t_\psi)$, paths analogous to Fig.~\ref{fig:pinchedGeometry}(b) arise wherein the incident energy splinters into displaced fermion wavepackets.
These paths underlie similar interference corrections to $\langle s^z_2(t)\rangle$ but with different sensitivity to bulk quasiparticles:
The $\mathcal{O}(t_\psi)$ interference correction $g^a_{\rm int}$ satisfies $g^{a =\mathds{1}}_{\rm int} = g^{a = \psi}_{\rm int} = -g^{a =\sigma}_{\rm int}$.  The first equality arises because a fermion encircling either a boson or fermion yields a trivial statistical phase; the minus sign in the second equality reflects the $-1$ statistical phase acquired when a fermion encircles an Ising anyon.  
Fermion tunneling thus shifts the late-time probabilities in Fig.~\ref{fig:pinchedGeometry}(c), but, importantly, all three quasiparticle types generically remain distinguishable.

{\bf \emph{Discussion.}}~Our theory implicitly assumes that the spin liquid's bulk excitation gap exceeds the ancillary-spin Zeeman energy so that the gapless edge mode provides the dominant energy-shuttling medium.  
Given the $\mathcal{O}(\unit[10]{T})$ field required to reach the purported non-Abelian spin liquid phase in $\alpha$-$\mathrm{RuCl}_3$, ancillary spins with suppressed, tunable $g$-factors (as in, e.g., Refs.~\onlinecite{Doty_2006, Prechtel2015,studenikin_2019}) are desirable both to satisfy this constraint and for probing the oscillatory interference correction in Eq.~\eqref{eq:Ptun}.
We stress, however, that physical spins may be substituted for arbitrary addressable quantum two-level systems that can locally couple to the edge.

At finite temperature edge-phonon coupling---which can be important for thermal transport \cite{Ye_2018, Vinkler_Aviv_2018}---allows injected energy to leak into the bulk, even when the spin-liquid gap is `large'.  
We argue that phonon-transport corrections play a minor role in our context for two reasons: $(i)$ Whereas the edge mode serves as an energy waveguide between the ancillary spins, energy carried by phonons has a relatively low probability of reaching the absorber spin ${\bf s}_2$.  
$(ii)$ Phonons that do reach ${\bf s}_2$ will generally not arrive simultaneously with edge wavepackets; timing considerations thus further suppress the impact of phonon-mediated energy shuttling.
Phonon leakage can still reduce the energy-absorption `kick' for well-timed pulses [Figs.~\ref{fig:unpinchedGeometry}(b) and \ref{fig:pinchedGeometry}(c)] but is not expected to qualitatively alter our conclusions.

Edge-state interrogation does not require phase coherence and thus provides an enticing preliminary experiment.
For time-domain anyon interferometry, by contrast, the edge distance $x_2-x_1$ between the ancillary spins must be sufficiently small that phase coherence persists.  
Moreover, the $\lambda_j(t)$ time scale $\tau$ should satisfy $h_z\tau \gg 1$ to maintain approximate energy conservation along with $L_{\rm perimeter} > v\tau$ ($L_{\rm perimeter}$ is the spin liquid's \emph{total} perimeter) to avoid trivial self-interference of pulses.
To get a very rough sense of scales, if $h_z \sim \unit[1]{K}$, $v \sim \unit[10^4]{m/s}$, and $L_{\rm perimeter} \sim \unit[10]{\mu m}$, then these conditions are satisfied for $\tau \sim \unit[1]{ns}$.
When $h_x = 0$ so interference comes strictly from Eq.~\eqref{eq:Ptun} we further desire that $v\tau \sim L_a$; with $L_a \sim \unit[1]{\mu m}$ this condition holds for $\tau \sim \unit[0.1]{ns}$.

We expect that temporal control over $\lambda_j(t)$ can be substituted by a time-dependent Zeeman energy (or other qubit splitting) as has been explored in the context of Levitons and FQH systems \cite{Keeling_2008, Wagner_2019}.
More generally, time-domain anyon interferometry naturally adapts to other chiral topological phases where electrical transport measurements are challenging (Mott insulators, cold atoms, etc.).  
The fractionalized excitations need not be non-Abelian; single-anyon detection requires only relative phase accumulation associated with braiding around a quasiparticle.

{\bf \emph{Acknowledgements.}}~We are grateful to Erik Henriksen, Jason Petta, Ady Stern, and Ashvin Vishwanath for illuminating discussions.
This work was supported by the Army Research Office under Grant Award W911NF17-1-0323;
	the National Science Foundation through grant DMR-1723367 and DMR-1848336;
	the Caltech Institute for Quantum Information and Matter, an NSF Physics Frontiers Center with support of the Gordon and Betty Moore Foundation through Grant GBMF1250;
	the Harvard-MIT CUA, ARO Grant W911NF-20-1-0163;
	the AFOSR-MURI Photonic Quantum Matter award FA95501610323;
	and the Walter Burke Institute for Theoretical Physics at Caltech.  
The final stage of this work was in part based on support by the U.S.~Department of Energy, Office of Science through the Quantum Science Center (QSC), a National Quantum Information Science Research Center.

\clearpage
\setcounter{secnumdepth}{2}
\begin{widetext}
\appendix
\section{Perturbative calculation of \texorpdfstring{$\langle s_2^z(t) \rangle$}{<s2z>}}
\label{app:PerturbS2Z}

This Appendix derives the perturbative expression of $\langle s_2^z(t) \rangle$ given in Eq.~\eqref{eq:sz2_expectation}.
To this end we turn off the transverse field $h_x$ acting on the emitter spin ${\bf s}_1$ since it does not affect the calculation, and organize the perturbative expansion by assuming
\begin{equation}
  \Lambda_j = \frac{\bar\lambda_j h_z}{v^2} \sim \frac{h_x}{h_z} \ll 1 ~{\rm and}~h_z \tau \gg 1.
  \label{perturbative_regime}
\end{equation}
Treating $V(t) = \sum_{j=1,2}\frac{\lambda_j(t)}{2\pi} s_j^x T(x_j)$ as a time-dependent perturbation, we expand the time-evolution operator in powers of $\lambda_j(t)$ via
\be
U(t) = U_0(t) -i\int_0^t \dif t_a U_0(t-t_a)V(t_a)U_0(t_a) -  \int_0^t \dif t_b \int_0^{t_b} \dif t_a U_0(t-t_b)V(t_b)U_0(t_b-t_a)V(t_a)U_0(t_a) + \cdots
\label{eq:dysonSeries}
\ee
with $U_0(t) = e^{-iH_0t}$ the evolution operator for the free Hamiltonian $H_0$ at zero spin-edge coupling.
Since the two spins and the edge all decouple in $H_0$, we can further decompose $U_0(t) = U_{s_1}(t)U_{s_2}(t)U_e(t)$, where $U_{s_1}(t) = \exp[-ith_z s_2^z]$, $U_{s_2}(t) = \exp\left[-it(h_x s_2^x + h_z s_2^z) \right]$, and $U_e(t) = \exp\left[-vt\int_x \gamma \partial_x \gamma \right]$ are the time-evolution operators for ${\bf s}_1$, ${\bf s}_2$, and the edge, respectively. 
In the spirit of our perturbative analysis we implicitly retain $U_{s_2}(t)$ only to $\mathcal{O}(h_x^2)$.
The time dependence of $\langle s_2^z(t) \rangle$ can be partitioned into three pieces that we examine sequentially below:
\begin{equation}
  \langle s_2^z(t) \rangle = S_{\rm precession}(t) + S_{\rm relax}(t) + S_{\rm shuttling}(t).
\end{equation}
Here $S_{\rm precession}$ describes standard spin precession from the Zeeman field, $S_{\rm relax}(t)$ describes processes where spin ${\bf s}_2$ flips via $h_x$ and then relaxes by shedding energy into the edge, and $S_{\rm shuttling}$ is the crucial term that captures energy shuttling from ${\bf s}_1$ to ${\bf s}_2$.
We will specifically evaluate the leading nontrivial contribution to each of these terms assuming the perturbative criteria in Eq.~\eqref{perturbative_regime}.
Henceforth in this Appendix we fix $v=1$ and then restore appropriate factors of $v$ by dimensional analysis.

\subsection{Precession}

At $\mathcal{O}(\bar\lambda_j^0)$ in the spin-edge couplings, ${\bf s}_2$ simply precesses due to $h_x \not= 0$.
This contribution is described by
\be
S_{\rm precession}(t) = -1 + \left(h_x / h_z\right)^2\sin^2(h_z t).
\ee

\subsection{Spin-edge relaxation}

Next we consider the correction $S_{\rm relax}(t)$, which depends on the $\lambda_2(t)$ coupling between ${\bf s}_2$ and the edge but \emph{not} the $\lambda_1(t)$ coupling between ${\bf s}_1$ and the edge.
Normal ordering in the stress-energy tensor $T(x_2)$ in the $\lambda_2(t)$ term precludes an $\mathcal{O}(\bar\lambda_2)$ contribution to $S_{\rm relax}(t)$.
The leading-order contribution then comes at $\mathcal{O}(\bar\lambda_2^2)$ from terms like $\langle \cdots s_2^z \cdots V \cdots V \cdots \rangle$ and $\langle \cdots V \cdots s_2^z  \cdots V \cdots \rangle$, where the ellipses denote appropriate $U_0$ factors.
To be more precise, denote the wavefunction by
\begin{equation}
  \ket{\phi(t)} = \ket{\lambda_2^0} + \ket{\lambda_2^1} + \ket{\lambda_2^2} + \cdots,
\end{equation}
where $\ket{\lambda_2^k}$ is the $\mathcal{O}(\bar \lambda_2^k)$ component (with $\bar \lambda_1 = 0$).
Then to $\mathcal{O}(\bar \lambda_2^2)$ we have
\begin{equation}
  S_{\rm relax}(t) = \bra{\lambda_2^1}s^z_2\ket{\lambda_2^1} + 2 \RePart\bra{\lambda_2^0} s^z_2 \ket{\lambda_2^2}.
  \label{Srelax_decomp}
\end{equation}
Evaluation is conveniently carried out using momentum-space edge fermion operators.
Fourier transforming via $\gamma(x) = \frac{1}{\sqrt{2}}\int \frac{\dif k}{2\pi}e^{ikx}\gamma_k$, the momentum operators obey $\gamma_k^\dagger = \gamma_{-k}$ along with the anti-commutation relation $\{\gamma_k, \gamma_q\} = 2\pi\delta(k+q)$.  
The relevant fermion bilinear in $T(x_2)$ may be written as
\be
\begin{aligned}
\norder{\gamma\partial_x\gamma} &= \frac12\int_0^\Lambda \frac{\dif k_1\, \dif k_2}{(2\pi)^2} (ik_2)\left[e^{i(k_1+k_2)x}\gamma_{k_1}\gamma_{k_2} - e^{-i(k_1+k_2)x}\gamma_{k_1}^\dagger\gamma_{k_2}^\dagger + e^{-i(k_1-k_2)x}\gamma_{k_1}^\dagger\gamma_{k_2} - e^{i(k_1-k_2)x}\norder{\gamma_{k_1}\gamma_{k_2}^\dagger}\right]\\
&= \frac{i}{2}\int_0^\Lambda \frac{\dif k_1\, \dif k_2}{(2\pi)^2} \left[e^{i(k_1+k_2)x}k_2\gamma_{k_1}\gamma_{k_2} - e^{-i(k_1+k_2)x}k_2\gamma_{k_1}^\dagger\gamma_{k_2}^\dagger + (k_2 + k_1)e^{-i(k_1-k_2)x}\gamma_{k_1}^\dagger\gamma_{k_2}\right]
\end{aligned}
\ee
with $\Lambda$ a momentum cutoff to be later taken to infinity.

Using this momentum-space representation, we obtain a first-order wavefunction correction
\be
\ket{\lambda_2^1} = i\int_0^t\dif t_a U_{s_2}(t-t_a)s_2^xU_{s_2}(t_a)\ket{s^z_2 = \dn} \int_0^\Lambda \frac{\dif k_1 \, \dif k_2}{(2\pi)^2} \frac{k_2\lambda_2(t_a)}{2}e^{-i(k_1+k_2)(x_2 + t-t_a)}\gamma_{k_1}^\dagger\gamma_{k_2}^\dagger\ket{0}.
\ee
(We have dropped the ${\bf s}_1$ subsystem since it evolves trivially here.)
The first term on the right side of Eq.~\eqref{Srelax_decomp} follows as
\begin{align}
\bra{\lambda_2^1}s^z_2\ket{\lambda_2^1} &= \int_0^t \dif t_a \dif t_b \int_0^\Lambda \frac{\dif k_1 \, \dif k_2}{(2\pi)^2} \frac{\lambda_2(t_a)\lambda_2(t_b)}{4}k_2(k_2-k_1)e^{-i(k_1+k_2)(t_b-t_a)}
\nonumber \\
&\times \bra{\dn} U_{s_2}^\dagger(t_b)s_2^xU_{s_2}^\dagger(t-t_b)s_2^zU_{s_2}(t-t_a)s_2^xU_{s_2}(t_a) \ket{\dn}.
\end{align}
The full time dependence is unwieldy to write out here, so let us just consider the late-time limit, $t \gg t_2$.
Due to the $\lambda_2(t)$ terms, the integrand only has appreciable weight where $t_a \approx t_b \approx t_2$; we are thus free to extend the $t_{a,b}$ integration bounds to $\pm \infty$.
Carrying out these integrals and dropping terms that will be exponentially suppressed in $h_z\tau$ for all $k_1$ and $k_2$ yields
\be
\bra{\lambda_2^1}s^z_2\ket{\lambda_2^1} = \int_0^\Lambda \frac{\dif k_1\, \dif k_2}{(2\pi)^2} \frac{\bar\lambda_2^2 k_2(k_1-k_2)\pi\tau^2}{8}\left(\frac{h_x}{h_z}\right)^2e^{-(2h_z+k_1+k_2)^2\tau^2}\left[e^{8h_z(k_1+k_2)\tau^2} + 4e^{4h_z(h_z+k_1+k_2)\tau^2}\right].
\ee
Performing the remaining momentum integrals and sending $\Lambda \rightarrow \infty$ gives the late-time correction
\be
\bra{\lambda_2^1}s^z_2\ket{\lambda_2^1} = -\frac{1}{24\sqrt{\pi}}\Lambda_2^2 \left(\frac{h_x}{h_z}\right)^2(h_z \tau) + \cdots \label{eq:sRelax1},~~~~~(t \gg t_2).
\ee
The ellipsis includes terms that are exponentially small in $h_z\tau$ (due to approximate energy conservation) as well as corrections that are down by powers of $1/(h_z\tau)$.

The second term on the right side of Eq.~\eqref{Srelax_decomp} arises from processes where ${\bm s}_2$ deposits some energy to the edge and then immediately re-absorbs it, and more explicitly reads
\begin{align}
2 \RePart\bra{\lambda_2^0} s^z_2 \ket{\lambda_2^2} &= 2\RePart\left[-\int_0^t \dif t_b \int_0^{t_b} \dif t_a \bra{s_2^z = \dn}\otimes\bra{0} U_0^\dagger(t) s_2^z U_0(t-t_b)V(t_b)U_0(t_b - t_a)V(t_a)U_0(t_a) \ket{0}\otimes\ket{s_2^z = \dn}\right]
\nonumber \\
&=2\RePart\biggl[-\int_0^t \dif t_b \int_0^{t_b} \dif t_a \int_0^\Lambda \frac{\dif k_1 \, \dif k_2}{(2\pi)^2} \bra{\dn}U_{s_2}^\dagger(t)s_2^zU_{s_2}(t-t_b)s_2^xU_{s_2}(t_b-t_a)s_2^xU_{s_2}(t_a)\ket{\dn}
\nonumber \\ 
&~~~~~~~~\times \frac{\lambda_2(t_a)\lambda_2(t_b)}{4}  k_2(k_1-k_2)e^{-i(k_1+k_2)(t_b-t_a)}\biggr].
\end{align}
We will again derive explicit results only for the late-time limit.
As before the integrand carries appreciable weight only for $t_a \approx t_b \approx t_2$.
Upon changing integration variables from $(t_b,t_a)$ to $(t_b,\delta t = t_b-t_a)$, this fact allows us to benignly integrate $t_b$ from $-\infty$ to $+\infty$ and integrate $\delta t$ from 0 to $+\infty$.  
Performing these time integrals, then integrating over $k_{1,2}$, and finally sending $\Lambda \rightarrow +\infty$ gives a late-time correction
\be
2 \RePart\bra{\lambda_2^0} s^z_2 \ket{\lambda_2^2}  = \frac{1}{24\sqrt{\pi}}\Lambda_2^2\left(\frac{h_x}{h_z}\right)^2(h_z\tau)[1-2\cos(2h_z t)] + \cdots,~~~~~(t\gg t_2).\label{eq:sRelax2}
\ee
Similar to Eq.~\eqref{eq:sRelax1}, we explicitly displayed only the leading terms in $h_z\tau$.

Combining Eqs.~\eqref{eq:sRelax1} and \eqref{eq:sRelax2} and restoring appropriate factors of $v$ yields an overall correction
\be
S_{\rm relax}(t \gg t_2) = -\frac{1}{12\sqrt{\pi}}\Lambda_2^2 \left(\frac{h_x}{h_z}\right)^2 \left(h_z \tau \right) \cos(2h_z  t). 
\label{eq:dyson_OLL}
\ee
The full time dependence requires a more nuanced treatment than we gave above.  
However, for wide pulses ($h_z\tau \gg 1$), this general time dependence is well-approximated by simply multiplying Eq.~\eqref{eq:dyson_OLL} by $\frac{1 + \erf\left[(t-t_2)/\tau\right]}{2}$---which just describes the correction turning on as the pulse $\lambda_2(t)$ passes.

\subsection{Energy shuttling}

We finally turn to the energy-shuttling term in Eq.~\eqref{eq:sz2_expectation}, focusing for now on the $h_x = 0$ limit.
Let $\ket{\lambda_1^1 \lambda_2^1}$ denote the $\mathcal{O}(\bar\lambda_1\bar\lambda_2)$ wavefunction correction.
Since both spins flip in $\ket{\lambda_1^1 \lambda_2^1}$ relative to the free-evolution wavefunction component corresponding to $\lambda_j = 0$ (and since we are taking $h_x = 0$), the leading energy-shuttling correction reads
\begin{equation}
  S_{\rm shuttling}(t) = \bra{\lambda_1^1 \lambda_2^1}s^z_2 \ket{\lambda_1^1 \lambda_2^1}.
\end{equation}
The wavefunction correction at time $t$ is given by 
\begin{align}
\ket{\lambda_1^1 \lambda_2^1} &= \int_0^t \dif t_b \int_0^{t_b} \dif t_a e^{2ih_z(t_b - t_a)}\lambda_1(t_a)\lambda_2(t_b) U_e(t-t_b)\norder{\gamma \partial_x \gamma}\rvert_{x_2} U_e(t_b-t_a) \norder{\gamma \partial_x \gamma}\rvert_{x_1}\ket{0}\ket{s_1^z=\dn, s_2^z = \up}.
\nonumber \\
& = \int_0^t \dif t_b \int_0^{t_b} \dif t_a \int_0^\Lambda \frac{\dif k_1 \dif k_2}{(2\pi)^2} \frac{\lambda_1(t_a)\lambda_2(t_b)}{4} k_2(k_1-k_2)e^{-i(k_1+k_2)(t_b - t_a + x_1 - x_2)}e^{i2h_z(t_b-t_a)}\ket{0}\ket{s_1^z = \dn, s_2^z = \up} 
\nonumber \\
&+ \cdots.
\end{align}
In the second line we explicitly displayed only the energy-shuttling component for which ${\bf s}_2$ absorbs both fermions injected by ${\bf s}_1$.
[Retaining terms lumped into the ellipsis generates only subleading contributions to $\langle s_2^z(t)\rangle$.]
Since the integrand carries appreciable weight only for well-separated times $t_b \approx t_2$ and $t_a \approx t_1$, at late times $t \gg t_2$ we can integrate both time integrals from $-\infty$ to $+\infty$.  
Invoking similar approximations as above then yields
\be
\ket{\lambda_1^1 \lambda_2^1} \approx -\frac{1}{6\sqrt{\pi}}e^{i2h_z(x_2 - x_1)}\left(\Lambda_1\Lambda_2\right)(h_z \tau)\exp\left[-\frac{\Delta^2}{(2\tau)^2}\right]\ket{0}\otimes\ket{s_1^z=\dn, s_2^z = \up} + \cdots,~~~~~(t\gg t_2).
\label{eq:s_shuttling_wf}
\ee
In Eq.~\eqref{eq:s_shuttling_wf} we expressed the prefactor of the exponential in the limit $|\Delta| \lesssim \tau$, since for larger $|\Delta|$ the correction is in any case negligible.
While the above holds only for late times, the solution at arbitrary time is well-approximated by introducing an error function that turns on at $t = t_2 -\Delta/2$, as we will see in the subsequent Appendix. 
Restoring $v$ factors, we arrive at the late-time correction 
\be
S_{\rm shuttling}(t \gg t_2) = \frac{1}{36\pi}\left(\Lambda_1\Lambda_2\right)^2 (h_z \tau)^2 \exp\left[-\frac{\Delta^2}{2(v\tau)^2}\right].
\label{Sshuttling_final}
\ee

Suppose that we now resurrect a non-zero transverse field $h_x$.  
For $h_x \not= 0$, the shuttling correction in Eq.~\eqref{eq:s_shuttling_wf} may overlap with the $\mathcal{O}(h_x^2)$ wavefunction correction in which both spins flip due to spin precession.
This wavefunction correction reads $-\left(\frac{h_x}{h_z}\right)^2\sin^2(h_zt)\ket{0}\otimes\ket{s_1^z = \dn, s_2^z = \up}$ and gives an additional energy-shuttling term 
\begin{align}
S_{\rm shuttling}^{h_x} &= -\left(\frac{h_x}{h_z}\right)^2\sin^2(h_zt)\bra{0}\otimes\bra{s_1^z = \dn, s_2^z = \up} s^z_2 \ket{\lambda_1^1 \lambda_2^1} + c.c.
\nonumber \\
&=\frac{1}{3\sqrt{\pi}}\sin^2(h_z t)\cos\left[\frac{2h_z(x_2 - x_1)}{v}\right] \left(\frac{h_x}{h_z}\right)^2\Lambda_2(h_z \tau)\exp\left[-\frac{\Delta^2}{(2v\tau)^2}\right].
\label{Sshuttling_hx}
\end{align}
Our perturbation hierarchy specified in Eq.~\eqref{perturbative_regime} implies that this piece is down by a factor of $1/(h_z\tau)$ compared to Eq.~\eqref{Sshuttling_final}. 
Nevertheless, Eq.~\eqref{Sshuttling_hx} is interesting in that it encodes oscillations in $h_z$ that may be observable if the ancillary spins have tunable Zeeman energies.

\section{Ising-anyon tunneling}
\label{app:IsingTunnel}

Here we give a careful CFT treatment of Ising-anyon tunneling, first explicitly deriving the $\mathcal{O}(\tilde t_\sigma)$ interference correction $f^{a = \mathds{1}}_{\rm int}$ to $\langle s_2^z(t) \rangle$ (taking $h_x = 0$ which suffices for the leading contribution) and then examining the manner in which incident energy partitions among the two Ising anyons that splinter at the constriction.
For convenience we fix $v=1$ throughout this Appendix and again restore the appropriate units at the end.
It is also useful to work with an alternately normalized fermion operator $\psi(x) = \sqrt{4\pi}\gamma(x)$ so that the stress-energy tensor $T = -2\pi i \norder{\gamma\partial_x\gamma} = -\frac{i}{2}\norder{\psi\partial_x\psi}$ exhibits correlations $\langle T(z_1)T(z_2)\rangle = \frac{1}{4z_{12}^4}$ with $z_{12} = z_1 - z_2$.

\subsection{Correction to \texorpdfstring{$\langle s_2^z(t) \rangle$}{<s2z>}}

The interference correction of interest reads
\begin{align}
  f_1^{a = \mathds{1}}(t) =2 \RePart\bra{\phi^{a=\mathds{1}}_0}s^z_2\ket{\phi^{a=\mathds{1}}_1}.
\end{align}
We will evaluate $f_1^{a = \mathds{1}}(t)$ for a more general problem with tunneling Hamiltonian 
\begin{equation}
  H_{\rm tun} = t_\varphi e^{-i\pi h_\varphi}\varphi(x_b)\varphi(x_t),
\end{equation}
where $\varphi$ denotes a primary field of the $c = 1/2$ CFT with conformal weight $h_\varphi$.  
Doing so allows us to attack in a unified manner the cases where Ising anyons \emph{or} fermions tunnel across the constriction, though we are primarily interested in Ising-anyon tunneling here.

Since the energy-shuttling component of $\ket{\phi^{a=\mathds{1}}_0}$ is proportional to $\ket{0}\otimes\ket{\dn,\up}$, as a first step we will extract the overlap 
\begin{align}
	\bra{0}\otimes\braket{\dn,\up | \phi_1^{a = \mathds{1}}} = \frac{i}{(2\pi)^2}\int_0^t \!\dif t_b \int_0^{t} \!\dif t_a \, \lambda_1(t_a)\lambda_2(t_b)e^{2ih_z(t_b-t_a)} \Braket{ T(t_b, x_2) \left[\int_{t_c} H_{\rm tun}(t_c) - \langle H_{\rm tun}(t_c)\rangle \right] T(t_a, x_1) }.
\label{eq:tunnelingExpr}
\end{align}
Subtraction of the vacuum expectation $\langle H_{\rm tun} \rangle$ accounts for the (unimportant) correction to the edge vacuum state $\ket{0}$ induced by anyon tunneling.
The correlator given in Eq.~\eqref{eq:tunnelingExpr} admits the explicit form
\begin{align} \begin{aligned}
	& \langle T(z_1)\varphi(\eta_1)\varphi(\eta_2)T(z_2)\rangle - \langle  T(z_1)T(z_2)\rangle \langle \varphi(\eta_1)\varphi(\eta_2) \rangle \\ &\qquad\qquad = \frac{2h_\varphi \eta_{12}^{2-2h_\varphi}}{z_{12}^2(z_1-\eta_1)(z_1-\eta_2)(z_2-\eta_1)(z_2-\eta_2)} + \frac{h_\varphi^2\eta_{12}^{4-2h_\varphi}}{(z_1-\eta_1)^2(z_1-\eta_2)^2(z_2-\eta_1)^2(z_2-\eta_2)^2},
\end{aligned} \label{eq:genericTTcorrelator} \end{align}
where we assume normalization such that $\langle \varphi(\eta_1)\varphi(\eta_2) \rangle = \eta_{12}^{-2h_\varphi}$.
Taking $z_1 = i(t_b - x_2) + \epsilon$, $z_2 = i(t_a - x_1) - \epsilon$, $\eta_1 = i(t_c - x_b)$, and $\eta_2 = i(t_c-x_t)$ in Eq.~\eqref{eq:genericTTcorrelator}, with $\epsilon\rightarrow 0^+$ an infinitesimal regularizing constant, we may integrate over $t_c$ to find
\begin{align} \begin{aligned}
	&\int_{t_c} \left[\frac{2h_\varphi \eta_{12}^{2-2h_\varphi}}{z_{12}^2(z_1-\eta_1)(z_1-\eta_2)(z_2-\eta_1)(z_2-\eta_2)} + \frac{h_\varphi^2\eta_{12}^{4-2h_\varphi}}{(z_1-\eta_1)^2(z_1-\eta_2)^2(z_2-\eta_1)^2(z_2-\eta_2)^2}\right]
	\\&\qquad\qquad \equiv 2\pi i e^{i\pi h_\varphi}L_a^{-3-2h_\varphi}j(y-i \epsilon).
	\label{eq:TTcorrelator_tauIntegral}
\end{aligned} \end{align}
In the second line we defined 
\begin{equation}
  y = \frac{t_b - t_a - L_x}{L_a},
\end{equation}
where $L_a = x_b - x_t$ as usual denotes the length of the enclosed region and $L_x = x_2 - x_1$ is the edge distance between the ancillary spins, along with the function
\begin{align}
	j(z) = 4h_\varphi\left[\frac{z^4 + (5h_\varphi - 2)z^2 + (1-h_\varphi)}{z^3(z+1)^3(z-1)^3}\right].
\end{align}
For brevity we also absorbed a constant factor $2/L_a$ into the regularizing constant $\epsilon$.

Additionally defining 
\begin{equation}
  f(t_b,y) = \lambda_1(t_a)\lambda_2(t_b)e^{2ih_z(t_b-t_a)}
  \label{fdef}
\end{equation}
(the $t_a$ dependence is implicit in $y$) allows us to compactly express Eq.~\eqref{eq:tunnelingExpr} as 
\begin{align}
  \bra{0}\otimes\braket{\dn,\up \vert \phi_1^{a = \mathds{1}}} &= -\frac{t_\varphi}{2\pi L_a^{3+2h_\varphi}} \int_0^t \dif t_b \int_0^{t} \dif t_a f(t_b,y) j(y-i \epsilon). 
\end{align}
For finite-width pulses $\lambda_j$, the integrand is negligible unless $t_b \approx t_2$ and $t_a \approx t_1$, which once again allows us to extend the integration limits:
\begin{align}
  \bra{0}\otimes\braket{\dn,\up \vert \phi_1^{a = \mathds{1}}} &\approx -\frac{t_\varphi}{2\pi L_a^{3+2h_\varphi}} \int_{-\infty}^t \dif t_b \int_{-\infty}^\infty \dif t_a f(t_b,y) j(y-i\epsilon)
= -\frac{t_\varphi}{2\pi L_a^{2+2h_\varphi}} \int_{-\infty}^t \dif t_b \int_{-\infty}^\infty \dif y f(t_b,y) j(y-i\epsilon).
\end{align}
Note that we have not yet taken the late-time limit.

To evaluate the $y$ integral it is helpful to first expand $j(z)$ as
\be
\begin{aligned}
j(z) =4 h_\varphi \biggl[ -\frac{2h_\varphi + 1}{z} - \frac{(1-h_\varphi)}{z^3} + \frac{2h_\varphi + 1}{2(z+1)} + \frac{h_\varphi}{(z+1)^2} + \frac{h_\varphi}{2(z+1)^3} + \frac{2h_\varphi + 1}{2(z-1)} - \frac{h_\varphi}{(z-1)^2} + \frac{h_\varphi}{2(z-1)^3} \biggr].
\end{aligned}\label{eq:tunnel_partialFrac_J}
\ee
We then need to evaluate integrals of the form
\be
\int \dif t_b \int \dif y f(t_b, y) \frac{1}{(y+c-i\epsilon)^n}
\ee
with $c = 0, \pm 1$ and $n = 1,2,3$.
Recalling that $\epsilon>0$, we employ a Schwinger parameterization of these integrals,
\be
\int \dif t_b \int \dif y f(t_b, y) \frac{1}{(y+c-i\epsilon)^n} = \int \dif t_b \int \dif y \int_0^\infty \dif a f(t_b, y) e^{-i a (y+c-i\epsilon)} \frac{i^n a^{n-1}}{\Gamma(n)},
\ee
which greatly facilitates integration with respect to $y$. 

We are left with the task of evaluating 
\begin{equation}
\bra{0}\otimes\braket{\dn,\up \vert \phi_1^{a = \mathds{1}}} =
-\frac{2 t_\varphi h_\varphi}{\pi L_a^{2+2h_\varphi}} \int_{-\infty}^t \dif t_b [F_0(t_b) + F_{-1}(t_b) + F_{1}(t_b)]
\end{equation}
with
\be
\begin{aligned}
F_0(t_b) &= \int_0^\infty \dif a \int_{-\infty}^{\infty} \dif y f(t_b, y) \exp\left[-ia(y-i\epsilon)\right] \left[-\frac{i(h_\varphi - 1)a^2}{2} -  i(2h_\varphi+1) \right] \\
F_{-1}(t_b) &= \int_0^\infty \dif a \int_{-\infty}^{\infty} \dif y f(t_b, y) \exp\left[-ia(y+1-i\epsilon)\right] \left[ -\frac{i h_\varphi a^2 }{4} - h_\varphi  a + \frac{i(1+2h_\varphi)}{2}  \right] \\
F_1(t_b) &= \int_0^\infty \dif a \int_{-\infty}^{\infty} \dif y f(t_b, y) \exp\left[-ia(y-1-i\epsilon)\right] \left[-\frac{i h_\varphi a^2}{4} +  h_\varphi a  + \frac{i(1+2h_\varphi)}{2}  \right].
\end{aligned}
\ee
Each line above corresponds to the contribution from a different path that energy may take to reach the downstream spin ${\bf s}_2$. Retaining the dominant ($n=3$) pieces only, we find
\begin{align}
\int_{-\infty}^t \dif t_b F_0(t_b) &\approx 2i\pi^{3/2}(1-h_\varphi) L_a^2  e^{2ih_zL_x} (\Lambda_1\Lambda_2)  \tau \exp\left[-\frac{\Delta^2}{(2\tau)^2}\right]\erfc\left[\frac{2(t_2-t) - \Delta}{2\tau}\right] \label{eq:F0_int} \\
\int_{-\infty}^t \dif t_b F_{-1}(t_b) &\approx -i \pi^{3/2} h_\varphi L_a^2 e^{2ih_z (L_x-L_a)} (\Lambda_1\Lambda_2) \tau \exp\left[-\frac{(\Delta + L_a)^2}{(2\tau)^2}\right]\erfc\left[\frac{2(t_2 - t) - L_a - \Delta}{2\tau}\right] \label{eq:Fn1_int}\\
\int_{-\infty}^t \dif t_b F_{1}(t_b) &\approx -i \pi^{3/2} h_\varphi L_a^2 e^{2ih_z (L_x+L_a)} (\Lambda_1\Lambda_2) \tau \exp\left[-\frac{(\Delta - L_a)^2}{(2\tau)^2}\right]\erfc\left[\frac{2(t_2 - t) + L_a - \Delta}{2\tau}\right] \label{eq:Fp1_int}.
\end{align}
Here the $\erfc$ terms correspond to a smooth ``turning-on'' of the correction as the pulses pass ${\bf s}_2$; the subleading terms that we suppress include additional time dependence that becomes relevant only when $h_z \tau$ is small. 
Collecting these results gives a late-time limit
\begin{align}
  \bra{0}\otimes\braket{\dn,\up \vert \phi_1^{a = \mathds{1}}} &= -i\sqrt{\pi}\frac{8t_\varphi h_\varphi}{L_a^{2h_\varphi}}(1-h_\varphi)e^{i 2h_z L_x}(\Lambda_1 \Lambda_2)\tau \label{eq:tunnel_shuttling}
  \nonumber \\
  &\times \left\{\exp\left[-\frac{\Delta^2}{(2\tau)^2}\right] - \frac{h_\varphi}{2(1-h_\varphi)}\left(e^{-2i h_z L_a}\exp\left[-\frac{(\Delta+L_a)^2}{(2\tau)^2}\right] + e^{2i h_z L_a}\exp\left[-\frac{(\Delta-L_a)^2}{(2\tau)^2}\right]\right)\right\}.
\end{align}


The overlap $\bra{0}\otimes\braket{\dn, \up \vert \phi_0^{a = \mathds{1}}}$ was already computed in Eq.~\eqref{eq:s_shuttling_wf} (for late times) via a momentum-space free-fermion calculation.  
As a consistency check on our formalism we reproduce this result using the CFT approach.  
We start by writing
\begin{equation}
  \bra{0}\otimes\braket{\dn, \up \vert \phi_0^{a = \mathds{1}}} = -\frac{1}{(2\pi)^2}\int_0^t \dif t_b \int_0^{t} \dif t_a \lambda_1(t_a)\lambda_2(t_b) e^{2ih_z(t_b-t_a)} \langle T(t_b,x_2)T(t_a,x_1)\rangle,
\end{equation}
which takes the same form as Eq.~\eqref{eq:tunnelingExpr} minus the tunneling term.
To parallel our treatment above, we rewrite this expression in terms of the coordinate $y$ defined previously and employ a Schwinger reparametrization, yielding
\be
\begin{aligned}
 \bra{0}\otimes\braket{\dn, \up \vert \phi_0^{a = \mathds{1}}} &\approx -\frac{1}{4(2\pi)^2}\int_{-\infty}^t \dif t_b \int_{-\infty}^{\infty} \dif t_a \lambda_1(t_a)\lambda_2(t_b) e^{2ih_z(t_b-t_a)}\frac{1}{(t_b-t_a-L_x - i \epsilon)^4} \\ 
& = -\frac{1}{4(2\pi)^2} \int_{-\infty}^t \dif t_b \int_{-\infty}^\infty \dif y f(t_b, y) L_a^{-3} (y-i\epsilon)^{-4} \\
& = -\frac{1}{4(2\pi)^2 L_a^3} \int_{-\infty}^\infty \dif t_b \int \dif y \int_0^\infty \dif a f(t_b, y) \exp\left[-ia(y-i\epsilon)\right] \frac{a^3}{6}.
\end{aligned}
\ee
Carrying out these integrals, we find a dominant contribution 
\be
\bra{0}\otimes\braket{\dn,\up \vert \phi_0^{a = \mathds{1}}} \approx -\frac{1}{12\sqrt{\pi}}e^{2ih_zL_x}(\Lambda_1 \Lambda_2)(h_z\tau) \erfc\left[\frac{2(t_2 - t) - \Delta}{2\tau}\right]\exp\left[-\frac{\Delta^2}{(2\tau)^2}\right].
\label{eq:trivialPath_TT}
\ee
This result is consistent with the analysis from the free-fermion calculation and, moreover, gives us more straightforward access to the full time dependence.

The interference term $f_1^{a=\mathds{1}}(t) = 2\RePart\braket{\phi_0^{a=\mathds{1}} \vert s_2^z \vert \phi_1^{a=\mathds{1}}}$ follows from Eq.~\eqref{eq:trivialPath_TT} and \eqref{eq:tunnel_shuttling}.
For late times we obtain
\be
\begin{aligned}
2\RePart\braket{\phi_0^{a=\mathds{1}} \vert s_2^z \vert \phi_1^{a=\mathds{1}}} &= -\frac{4t_\varphi h_\varphi^2}{3 L_a^{2h_\varphi}} (h_z\tau)(\Lambda_1 \Lambda_2)^2 \tau \sin(2h_z L_a) \exp\left[-\frac{\Delta^2}{(2\tau)^2}\right] \\ &\times \left(\exp\left[-\frac{(\Delta+L_a)^2}{(2\tau)^2}\right] - \exp\left[-\frac{(\Delta-L_a)^2}{(2\tau)^2}\right] \right).
\label{interference_correction}
\end{aligned}
\ee
The full time dependence arises upon simply restoring appropriate error-function terms.
Fixing $\varphi = \sigma$ and $h_\sigma = 1/16$ and resurrecting factors of $v$ by dimensional analysis, we then arrive at the late-time correction presented in Eq.~\eqref{eq:Ptun}.


In the previous Appendix we noted that the trivial shuttling path overlaps with the wavefunction component generated by $h_x$-induced precession of both spins [recall Eq.~\eqref{Sshuttling_hx}].  The $\mathcal{O}(\tilde t_\sigma)$ energy-shuttling paths similarly interferes with $h_x$-induced precession processes, generating an additional interference correction to $\langle s_2^z(t)\rangle$:
\begin{align}
&-\left(\frac{h_x}{h_z}\right)^2\sin^2(h_zt) 2\RePart \bra{0}\otimes\bra{s_1^z = \dn, s_2^z = \up} s^z_2  \ket{\phi_1^{a=\mathds{1}}} 
\nonumber \\
&~~~~~~~~~~= -\frac{16\sqrt{\pi}t_\varphi h_\varphi}{L_a^{2h_\varphi}}(\Lambda_1\Lambda_2)\tau\left(\frac{h_x}{h_z}\right)^2\sin^2(h_z t) 
\biggl\{(1-h_\varphi)\sin(2h_zL_x)\exp\left[-\frac{\Delta^2}{(2\tau)^2}\right] 
\nonumber \\
&~~~~~~~~~~ + \frac{h_\varphi}{2}\sin[2h_z(L_x-L_a)]\exp\left[-\frac{(\Delta+L_a)^2}{(2\tau)^2}\right] + \frac{h_\varphi}{2}\sin[2h_z(L_z+L_a)]\exp\left[-\frac{(\Delta-L_a)^2}{(2\tau)^2}\right]\biggr\}.
\end{align}
Similar to Eq.~\eqref{Sshuttling_hx}, this correction is smaller by a factor of $1/(h_z\tau)$ compared to Eq.~\eqref{interference_correction}, but has the virtue that it features additional oscillatory dependence on $h_z$ [via $\sin^2(h_z t)$] that may be probed with tunable Zeeman energies.  


\subsection{Energy-domain analysis}
\label{app:TunnelingEnergyDomain}

Since we work in the regime where energy is approximately conserved, it is interesting to now re-examine the preceding calculations, but starting in the energy domain.
In what follows we will closely follow the related analysis conducted in Ref.~\onlinecite{aasen2020}.
Consider an incoming state $\ket{T_\omega} = \int \frac{\dif t}{2\pi} e^{-i\omega t} T(t)\ket{0}$ with well defined energy $\omega$ and normalization $\braket{T_{\omega_1} \vert T_{\omega_2}} = \frac{c}{12}\omega_1^3 \delta(\omega_1 - \omega_2)$, where
$c=1/2$ is the central charge.
We will examine the first-order correction $\mathcal{A}_1(\omega; x_2, x_1) \equiv \mathcal{A}_1(\omega)$ to the transmission amplitude due to quasiparticle tunneling at the constriction, which satisfies
\be
\mathcal{A}_1(\omega)\delta(\omega - \omega') = -i\frac{12}{c\omega^3}\left \langle T_{\omega'}(x_2) \left [ \int_{t_c} H_\text{tun}(t_c) - \langle H_\text{tun}(t_c)\rangle \right] T_\omega(x_1) \right \rangle.
\label{A1def}
\ee
Upon taking the Fourier transform of the incoming and outgoing states we get
\be
\mathcal{A}_1(\omega)\delta(\omega - \omega') = -i \frac{12 t_\varphi}{c\omega^3} e^{-i\pi h_\varphi} \int \frac{\dif t_a \dif t_b}{(2\pi)^2} e^{i\omega' t_b} e^{-i\omega t_a} \left\langle T(t_b, x_2) \left[ \int_{t_c} H_\text{tun}(t_c) - \langle H_\text{tun}(t_c)\rangle \right] T(t_a, x_1) \right\rangle.
\ee
Next we pull out the phase factor set by the end points and constriction length $L_a$ to obtain
\be
\mathcal{A}_1(\omega)\delta(\omega - \omega') = -i \frac{12 t_\varphi}{c\omega^3} e^{-i\pi h_\varphi} e^{i\omega(L_x - L_a)} \int \frac{\dif t_a \dif t_b}{(2\pi)^2} e^{i\omega' t_b} e^{-i\omega t_a} \left\langle T(t_b) \left[ \int_{t_c} H_\text{tun}(t_c) - \langle H_\text{tun}(t_c)\rangle \right] T(t_a) \right\rangle.
\ee
The correlator can be rewritten as in Eq.~\eqref{eq:genericTTcorrelator}, now with $z_1 = i(t_b - L_a) + \epsilon$, $z_2 = it_a - \epsilon$, $\eta_1 = i(t_c - L_a)$, and $\eta_2 = it_c$.
Upon integrating over $t_c$ we obtain a result analogous to Eq.~\eqref{eq:TTcorrelator_tauIntegral},
\be
\mathcal{A}_1(\omega)\delta(\omega - \omega') = \frac{24  \pi t_\varphi}{c\omega^3}  L_a^{-3-2h_\varphi} e^{i\omega(L_x - L_a)}\int \frac{\dif t_a \dif t_b}{(2\pi)^2} j(y-i\epsilon) e^{i\omega' t_b} e^{-i\omega t_a},
\ee
with $j(z)$ as defined previously but where now $y = \frac{t_b - t_a - L_a}{L_a}$.
Changing coordinates and integrating over $t_a$ yields
\be
\mathcal{A}_1(\omega)\delta(\omega - \omega') = \frac{24\pi t_\varphi}{c\omega^3} L_a^{-2-2h_\varphi}e^{i\omega(L_x - L_a)}  \int \frac{\dif y}{2\pi} j(y-i\epsilon) e^{i\omega' L_a (y+1)} \delta(\omega - \omega').
\ee
What remains then is to evaluate
\be
\mathcal{A}_1(\omega) = \frac{24\pi t_\varphi}{c\omega^3} L_a^{-2-2h_\varphi} e^{i\omega L_x} \int \frac{\dif y}{2\pi} j(y-i\epsilon) e^{i\omega L_a y}.\label{eq:A1_jIntegral}
\ee
For the physically relevant $\omega\geq0$ regime, we can close the $y$-integration contour in the upper-half plane and arrive at
\be
\begin{aligned}
\mathcal{A}_1(\omega\geq 0) = -i \frac{48 \pi t_\varphi}{c\omega^3} L_a^{-2-2h_\varphi} h_\varphi e^{i\omega L_x } \biggl\{& \cos(\omega L_a) [-2 + h_\varphi(\omega^2L_a^2 - 4)] - 4h_\varphi \omega L_a\sin(\omega L_a) \\ & + 2(1 + 2h_\varphi) + \omega^2L_a^2(h_\varphi - 1)  \biggr\}.
\end{aligned}\label{eq:generic_A1}
\ee
Figure~\ref{fig:energy_partition}(a) plots the amplitude correction versus $\omega L_a$ for the case of Ising-anyon tunneling.

{\bf \emph{Energy partitioning.}}~When quasiparticles tunnel across the constriction, they carry some fraction of the incident energy.
The distribution of energies carried by tunneling quasiparticles may be more formally characterized by inverting the transmission-amplitude correction in Eq.~\eqref{eq:generic_A1}.
In particular we will find $f_\omega(\omega')$, the distribution of tunneled energy $\omega'$ given an incident energy $\omega$, such that 
\begin{equation}
  \mathcal{A}_1(\omega;L_a) = \int_{-\infty}^\infty \dif \omega' e^{i \omega' L_a} f_\omega(\omega').
  \label{f_def}
\end{equation}
On the left side we explicitly noted the $L_a$ dependence in $\mathcal{A}_1$ to emphasize that $f_\omega(\omega')$ does \emph{not} depend on $L_a$.
Furthermore we drop the trivial phase $e^{i\omega L_x}$ fixed by the end points since this is unimportant for the energy partitioning.
Here a negative value of $\omega'$ will not correspond physically to a negative frequency but rather to a quasiparticle tunneling across the constriction in the reverse direction.
The two tunneling directions could be separated by requiring $\omega' > 0$ and then treating separately $\pm L_a$, but by allowing $\omega' < 0$ to encode one of these paths we are able to calculate both in one fell swoop.

We would like to extract $f_\omega(\omega')$ from $\mathcal{A}_1$ by inverting Eq.~\eqref{f_def}.
Although states with negative frequency do not exist in the CFT, we now formally extend the domain of $\mathcal{A}_1$ defined through Eq.~\eqref{eq:A1_jIntegral} to negative $\omega$.  
In that frequency regime we can close the integration contour in Eq.~\eqref{eq:A1_jIntegral} in the lower-half plane, yielding $\mathcal{A}_1(\omega<0; L_a) = 0$ since in our regularization all poles in $j(y-i\epsilon)$ reside in the upper-half plane.  
In conjunction with Eq.~\eqref{eq:A1_jIntegral}, this continuation allows us to write
\begin{equation}
  \mathcal{A}_1(\omega;L_a) = \int_{-\infty}^\infty \dif \omega' e^{i \omega'L_a } f_\omega(\omega') = \frac{12 t_\varphi}{c\omega^4} L_a^{-2-2h_\varphi} \int_{-\infty}^\infty \dif \omega'  e^{i\omega' L_a}j\left( \frac{\omega'}{\omega} - i\epsilon\right) .\label{eq:continuing_A1}
\end{equation}
Observe next that, according to Eq.~\eqref{eq:tunnel_partialFrac_J}, the real part of $j(y-i\epsilon)$ is even in $y$ while the imaginary part is odd.  
Using this property and $\mathcal{A}_1(\omega<0; L_a) = 0$ allows us to replace $j \rightarrow 2i \ImPart j$ in the expression for the physically relevant amplitude with non-negative frequencies:
\begin{equation}
  \mathcal{A}_1(\omega\geq0;L_a) = \int_{-\infty}^\infty \dif \omega' e^{i \omega'L_a } f_\omega(\omega') = \frac{24 t_\varphi}{c\omega^4} L_a^{-2-2h_\varphi} \int_{-\infty}^\infty \dif \omega'  e^{i\omega' L_a} i \ImPart j\left( \frac{\omega'}{\omega} - i\epsilon\right) .
  \label{A1_again}
\end{equation}
The form on the right side is particularly convenient since the imaginary component of $j$ features Dirac delta functions and derivatives thereof:
\be
\begin{aligned}
\ImPart j(y-i \epsilon) = \pi h_\varphi \biggl[& 4(2h_\varphi + 1)\delta(y) + 2(1-h_\varphi)\delta''(y) 
\\ & -2(2h_\varphi + 1)\delta(y+1) + 4h_\varphi\delta'(y+1) - h_\varphi\delta''(y+1)
\\ & -2(2h_\varphi + 1)\delta(y-1) - 4h_\varphi\delta'(y-1) - h_\varphi\delta''(y-1) \biggr].
\end{aligned}
\label{eq:tunnel_imJ}
\ee

Equation~\eqref{A1_again} suggests a form for $f_\omega(\omega')$, but it is important to recall that this function is $L_a$ independent by definition.  
The problematic $L_a$ dependence in front of the integral on the right side of Eq.~\eqref{A1_again} can be eliminated by repeated integration by parts.
Inversion of the transmission amplitude then yields the desired distribution function
\be
f_\omega(\omega') = -i\frac{t_\varphi}{\omega^{2-2h_\varphi}} \frac{24}{c} e^{-i\pi h_\varphi} \ImPart j^{(-2-2h_\varphi)}\left(\frac{\omega'}{\omega} - i\epsilon\right),
\label{fomega}
\ee
where $\ImPart j^{(-2-2h_\varphi)}$ is the $(2+2h_\varphi)^\text{th}$ integral of $\ImPart j$ defined in Eq.~\eqref{eq:tunnel_imJ}.
In general the $(s)^\text{th}$ integral of the Dirac delta function is given by $\delta^{(-s)}(x) = \Theta(x) x^{s-1}/\Gamma(s)$, where $\Theta(x)$ is the Heaviside step function.
Since $2+2h_\varphi > 2$ for primary fields with $h_\varphi > 0$, taking the $(2+2h_\varphi)^\text{th}$ integral of all the terms in Eq.~\eqref{eq:tunnel_imJ} remains well behaved.
This procedure then yields the general result
\be
\begin{aligned}
f_\omega(\omega') = -i\frac{t_\varphi h_\varphi}{\omega^3} \frac{96\pi}{\Gamma(1+2h_\varphi)} e^{-i\pi h_\varphi}
	\biggl\{& 2\left(\omega'\right)^{-1+2h_\varphi}\left[(\omega')^2 - h_\varphi(h_\varphi - 1)\omega^2\right]\Theta(\omega') \\
	& - \left(\omega' + \omega\right)^{-1 + 2h_\varphi}\left[\omega' + (1-h_\varphi)\omega\right]^2\Theta(\omega' + \omega) \\
	& - \left(\omega' - \omega\right)^{-1 + 2h_\varphi}\left[\omega' - (1-h_\varphi)\omega\right]^2\Theta(\omega' - \omega) \biggr\}.
\end{aligned}\label{eq:GenericPartition}
\ee
The solution at $\omega' < 0$ vanishes when $\omega' < -\omega$.
Recall that for $\omega' < 0$ the distribution function describes tunneling from the top edge with frequency $\abs{\omega'}$.
This cutoff at $\omega' = -\omega$ then reflects the physical constraint that tunneling from the top edge cannot carry more than the incident energy $\omega$ across the constriction.
Tunneling from the bottom edge, however, may carry arbitrarily large momentum, witnessed by $f_\omega(\omega' > \omega)$ not strictly vanishing in the above.

The distribution specified by Eq.~\eqref{eq:GenericPartition} features divergences at $\omega' = 0^+$, $\omega^+,$ and $(-\omega)^+$ with an exponent $-1+2h_\varphi$.
These divergences can be traced to singularities in $j\big(y = \frac{t_b-t_a}{L_a}\big)$ at $t_b-t_a = 0, \pm L_a$---and are thus associated with the three possible paths along which energy can reach ${\bf s}_2$ after the edge packet has fractionalized.
Comparing the magnitude of the terms in Eq.~\eqref{eq:GenericPartition} near their divergence, we observe that the divergences at $\omega'=(-\omega)^+$, $\omega^+$ are a factor of $\frac{h_\varphi}{2(1-h_\varphi)}$ smaller than the divergence at $\omega' = 0^+$.
This factor corresponds precisely to the ratios of the delayed/advanced pulses relative to the central pulse in the time domain [Eqs.~\eqref{eq:Fp1_int} and \eqref{eq:Fn1_int} versus Eq.~\eqref{eq:F0_int}].
In physical terms, these divergences and their relative strength indicate that an Ising anyon tunneling across the constriction from above preferentially carries nearly all the incident energy, whereas an Ising anyon tunneling from below predominantly (by a constant factor) carries negligible energy and secondarily carries slightly more than the full incident energy.
These tunneling propensities are advantageous since in our scheme flipping the downstream spin ${\bf s}_2$ requires that either essentially all or none of the energy is carried across the constriction.

For Ising anyons ($\varphi = \sigma, \, h_\sigma = 1/16$) Eq.~\eqref{eq:GenericPartition} becomes
\begin{align} \begin{aligned}
	f_\omega^{\varphi = \sigma}(\omega') = -i\frac{t_\varphi}{\omega^3}\frac{6\pi}{\Gamma(9/8)} e^{-i\pi / 16} \biggl\{& 2(\omega')^{-7/8} \Big[(\omega')^2 + \tfrac{15}{256}\omega^2\Big] \Theta(\omega') \\
	& - (\omega' + \omega)^{-7/8} \big(\omega' + \tfrac{15}{16}\omega\big)^2 \Theta(\omega'+\omega) \\
	& - (\omega' - \omega)^{-7/8} \big(\omega' - \tfrac{15}{16}\omega\big)^2 \Theta(\omega'-\omega) \biggr\}.
\end{aligned} \label{eq:IsingEnergyPartition} \end{align}
With fermion tunneling ($\varphi = \psi$, $h_\psi = 1/2$) the result instead reads
\be
f_\omega^{\varphi = \psi}(\omega') = -\frac{48 t_\varphi \pi}{\omega^{3}} \times\begin{cases} 0 & \abs{\omega'} >\omega,\\ \left(\omega'-\frac12\omega\right)^2 & 0 < \omega' < \omega,\\ \left(\omega' + \frac12\omega\right)^2 & -\omega < \omega' < 0. \end{cases}
\ee
Contrary to Ising-anyon tunneling, the energy-partitioning function for fermion tunneling features no divergences and vanishes outside of the range $\omega' \in [-\omega, \omega]$.
Furthermore, fermin tunneling from the upper edge (negative $\omega'$) is identical to fermion tunneling from the lower edge.
Figures~\ref{fig:energy_partition}(b) and (c) plot these distribution functions as a function of $\omega'/\omega$.

\begin{figure}
	\centering
	\includegraphics[width=.75\textwidth]{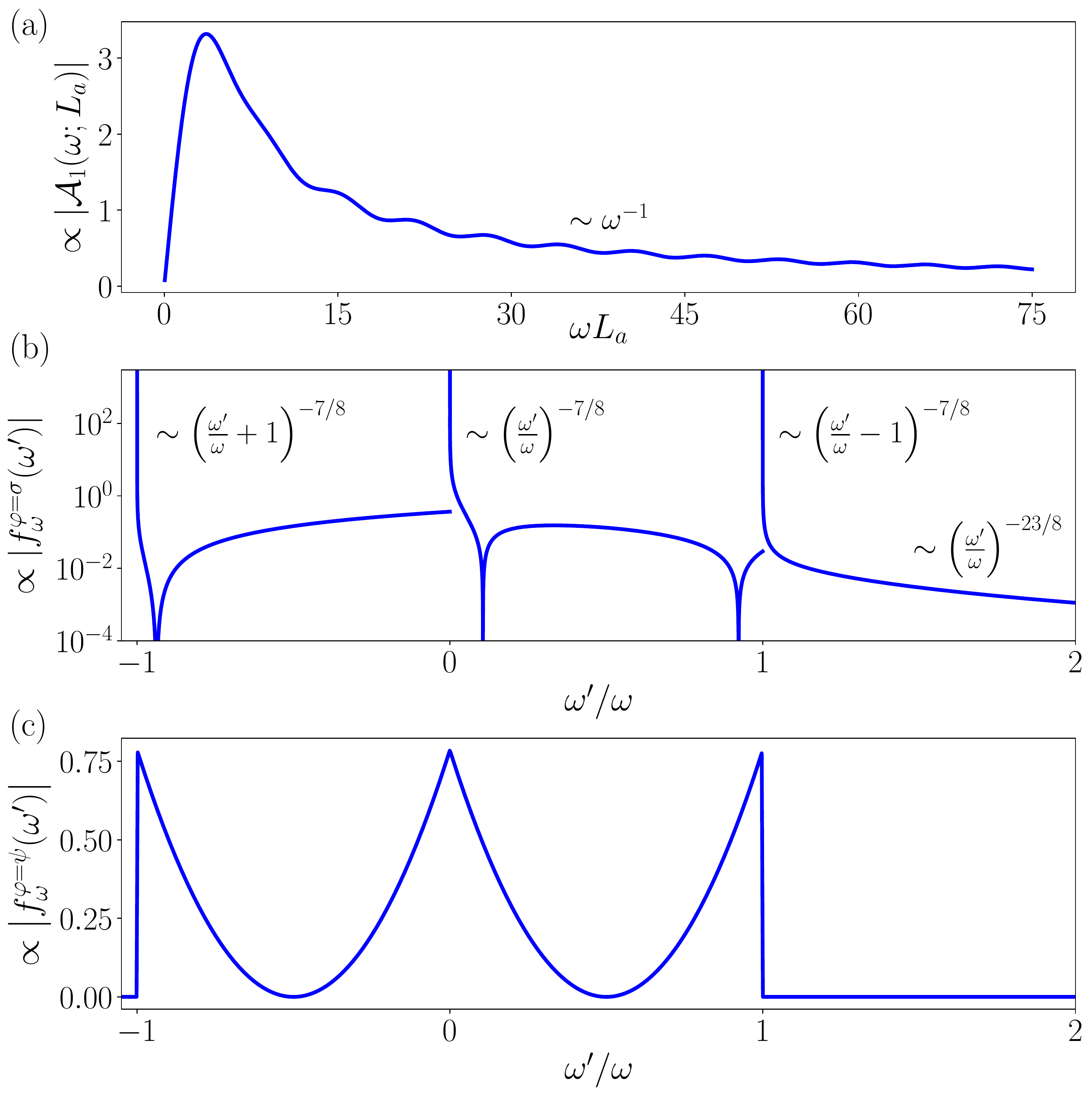}
	\caption{
	(a) First-order correction (in Ising-anyon tunneling $t_\sigma$) to the transmission amplitude in the energy domain. 
	For $\omega L_a \gg 1$, this tunneling correction scales as $1/\omega$.
	(b) Scaling behavior of the energy-partitioning distribution for Ising-anyon tunneling. 
	The distribution diverges at $\omega' = -\omega^+$, $0^+$, and $\omega^+$ with an exponent $-1+2h_\sigma = -7/8$.
	For $\omega' \gg \omega$ the distribution falls off like $(\omega')^{-23/8}$.
	(c) The corresponding energy-partitioning distribution for fermion tunneling features no divergences and vanishes where $\abs{\omega'} > \omega$.
	In (b) and (c), the $\omega'<0$ regime corresponds to processes where an Ising anyon tunnels across the constriction from above; there the energy carried by the tunneled Ising anyon is $|\omega'|$.
	}
	\label{fig:energy_partition}
\end{figure}

{\bf \emph{Relation to the time-domain calculation.}}~Equation~\eqref{A1def} allows us to alternatively express Eq.~\eqref{eq:tunnelingExpr} as
\be
\begin{aligned}
\bra{0}\otimes\braket{\dn,\up | \phi_1^{a = \mathds{1}}} & =\int_{t_b} \int_{t_a} \int_\omega \lambda_1(t_a)\lambda_2(t_b)e^{2ih_z(t_b-t_a)}e^{-i\omega(t_b - t_a)} \mathcal{A}_1(\omega; L_a)\frac{c\omega^3}{12} \\ 
& = \int_{t_b} \int_{t_a} \int_\omega \lambda_1(t_a)\lambda_2(t_b)e^{2ih_z(t_b-t_a)}e^{-i\omega(t_b - t_a - L_x)} \int_{\omega'} e^{i\omega' L_a} f_\omega(\omega') \frac{c\omega^3}{12}\\
& = \int_{t_b} \int_{t_a} \int_\omega \lambda_1(t_a)\lambda_2(t_b)e^{2ih_z(t_b-t_a)}e^{-i\omega(t_b - t_a - L_x)} \int_{\omega'} e^{i\omega' L_a} 2 t_\varphi \omega^{1+2h_\varphi} e^{-i\pi h_\varphi} j^{(-2-2h_\varphi)}\left(\frac{\omega'}{\omega}\right).\\
\end{aligned}
\ee
Carrying out these integrals yields results consistent with our time-domain analysis.
We will simply observe that first carrying out the integral over $t_a$ gives a Gaussian factor $\exp\left[-(2h_z - \omega)^2\tau^2/2\right]$ that enforces the energy-matching condition $\omega \rightarrow 2h_z$ as the pulse widths become arbitrarily large.
In the limit $\omega L_a \gg 1$ we have that $\abs{\mathcal{A}_1(\omega; L_a)} \propto 1/\omega$, which is responsible for the relative factor of $v/(L_a h_z)$ between the trivial path energy shuttling [Eq.~\eqref{eq:trivialPath_TT}] and the tunneling correction [Eq.~\eqref{eq:tunnel_shuttling}].
The slight smearing of $\omega$ over a narrow window about $2h_z$ partially smooths the divergences in $f_\omega(\omega')$, but the correction is still dominated by the case where approximately all or none of the energy crosses the constriction.

\end{widetext}

\bibliography{rucl3}

\end{document}